\documentclass{aa}
\usepackage[varg]{txfonts}
\usepackage{natbib}
\usepackage{graphicx}
\usepackage{times}
\usepackage{color}
\bibpunct{(}{)}{;}{a}{}{,}

\newcommand{\va}{V1624~Cyg }
\newcommand{\ve}{V1624~Cyg}

\newcommand{\respefo}{{\tt reSPEFO} }
\newcommand{\respefoe}{{\tt reSPEFO}}

\newcommand{\fotel}{{\tt FOTEL} }
\newcommand{\fotele}{{\tt FOTEL}}

\newcommand{\tria}{\hbox{$\bigtriangleup$}}
\newcommand{\ubv}{\hbox{$U\!B{}V$}}

\newcommand{\bv}{\hbox{$B\!-\!V$}}
\newcommand{\ub}{\hbox{$U\!-\!B$}}

\newcommand{\ubvr}{\hbox{$U\!B{}V\!R$}}
\newcommand{\ubvri}{\hbox{$U\!B{}V\!R\!I$}}
\newcommand{\uvby}{\hbox{$uvby$}}
\newcommand{\hp}{\hbox{$H_{\rm p}$}}
\newcommand{\oc}{\hbox{$O\!-\!C$}}

\newcommand{\arcm}{$^\prime$}

\newcommand{\m}{$^{\rm m}\!\!.$}

\newcommand{\ks}{km~s$^{-1}$}

\newcommand{\tef}{$T_{\rm eff}$ }

\newcommand{\ms}{M$_{\odot}$}
\newcommand{\rs}{R$_{\odot}$}
\newcommand{\cd}{c$\,$d$^{-1}$}

\newcommand{\ha}{H$\alpha$ }

\newcommand{\hb}{H$\beta$ }
\newcommand{\hg}{H$\gamma$ }
\newcommand{\hae}{H$\alpha$}

\newcommand{\he}{\ion{He}{i}~6678 }

\begin{document}

   \title{Spectroscopic orbit and variability of the Be star\\
                       V1624~Cyg = 28~Cyg
\thanks{Based on new spectroscopic observations
from the CCD coud\'e spectrograph of the Dominion Astrophysical Observatory;
on the CCD coud\'e spectra from the Ond\v{r}ejov Observatory,
and amateur spectra from the BeSS database.}
\fnmsep\thanks{Tables 3, 4, and 6 are available only
 at the CDS via anonymous ftp to cdarc.u-strasbg.fr (130.79.128.5)
 or via http://cdsweb.u-strasbg.fr/cgi-bin/qcat?J/A+A/}
}
  \author{P.~Harmanec\inst{1}\and
          S.~Yang\inst{2}\and
          P.~Koubsk\'y\inst{3}\and
          J.~Labadie-Bartz\inst{4}\and
          P.~Dole\v{z}al\inst{1}\and
          S.~Ranguin\inst{4}\and
          H.~Bo\v{z}i\'c\inst{5}\and
          J.~\v{S}vr\v{c}kov\'a\inst{1}\and
          M.~Zummer\inst{1}\and
          P.~Zasche\inst{1}\and
          H.~Ak\inst{6}
}

   \offprints{P. Harmanec\,\\
   \email \ Petr.Harmanec@matfyz.cuni.cz}

  \institute{
   Astronomical Institute of Charles University,
   Faculty of Mathematics and Physics,\hfill\break
   V~Hole\v{s}ovi\v{c}k\'ach~2, CZ-180 00 Praha~8 - Troja, Czech Republic
 \and
   Physics \& Astronomy Department, University of Victoria,
  PO Box 3055 STN CSC, Victoria, BC, V8W 3P6, Canada
 \and
   Astronomical Institute, Czech Academy of Sciences,
   CZ-251 65 Ond\v{r}ejov, Czech Republic
\and
LIRA, Observatoire de Paris, Universit\'e PSL, CNRS, Sorbonne Universit\'e,
Universit\'e Paris Cit\'e, CY Cergy Paris Universit\'e,
5~place Jules Janssen, 92195~Meudon, France
\and
  Hvar Observatory, Faculty of Geodesy, University of Zagreb,
  Ka\v{c}i\'ceva~26, 10000~Zagreb, Croatia
\and
  Astronomy and Space Sci. Dept., Science Faculty,
  Erciyes University, 38039, Kayseri, Turkiye
}
\date{Received \today, accepted May 19, 2025}

\abstract{
In recent years the idea, first formulated many decades ago, that the Be phenomenon
could be causally related to the duplicity of Be stars, has been repeatedly reconsidered
from various perspectives. It is important, therefore, to have reliable information
on Be stars, which are confirmed members of binary systems.
This study is devoted to V1624 Cyg = 28 Cygni, which was recently identified as
a binary with a compact secondary. By measuring the radial velocities (RVs) of the wings of
the H alpha emission line and using archival data and published RVs from the International
Ultraviolet Explorer, we demonstrate that the Be primary moves in the 359\fd26 orbit 
found recently from interferometry. Our preliminary radial-velocity solution leads to binary
masses of 5.6~\ms, and 0.66 ~\ms. Moreover, we documented large and irregular
spectral, brightness, and colour changes over a time interval of several decades
to show that the object never completely lost its circumstellar matter.
}


\authorrunning{P. Harmanec et al.}
\titlerunning{Spectroscopic orbit of the Be star \va}
\maketitle

\section {Introduction}
The bright B2-B3e star \va (also known as 28~Cyg, HD~191610, HR~7708,
BD+36$^\circ$3907, HIP~99303, and MWC 329;
$\alpha_{2000.0}$ = 20$^{\rm h}$09$^{\rm m}$25.$\!\!^{\rm s}$619,
$\delta_{2000.0}$ = +36$^\circ$50\arcm22\farcs64) has been
the~subject of numerous studies, especially since it was discovered to exhibit rapid 
light and line-profile changes. We refer readers to the
paper by \citet{hvar5}, where the first of these studies are summarized
in detail, and to the very detailed investigation by \citet{baade2018},
which is mainly based on space photometries and amateur \ha spectra from
the Be Star Spectra (BeSS) database \citep{neiner2011}. Hvar \ubv\ photometry reported
by \citet{hvar5} shows the range of rapid $V$ magnitude variations
from about 4\m9 to 5\m0 over the time interval from
JD~2446235 to 2447791, with no obvious secular change, and mild
long-term changes in both \bv\ and \ub\ indices. \citet{baade2018}
report long-term changes in the emission-line strength.
\citet{curtiss25} reported cyclic variations of the emission with a cycle
of 1373 days, which later faded away. These studies show that, like
many other early-type Be stars, \va exhibits a~complex set of variabilities,
both rooted in the star itself and its circumstellar disc.

\citet{losh32} analysed a collection of 205 photographic spectra
 from Ann Arbor Observatory and measured Balmer \hb and \hg
 lines. She published the mean radial velocities (RVs) of the absorption-line cores 
and the emission-line wings. She concluded that the star is probably a spectroscopic binary
 with a~226\fd0 orbital period and an~orbit with an eccentricity of 0.746 and
 a semi-amplitude of 12.2~\ks. She noted, however,
 that the RVs are also modulated by two other periodicities of 45\fd33, and
 1\fd51775.  Moreover, she further studied the changes in the emission strength
and also noted long-term RV changes in the early observations. For that reason,
she excluded these early RVs from the period analysis and tabulated them
separately. Her period analysis is thus based on 189 RVs.
In a study aimed to establish a fraction of binaries among bright
B and Be stars, \citet{abt78} concluded that \va has a constant radial velocity
(RV) of $-36.9\pm8.4$~\ks. They published 25 individual RVs
covering a~time interval over 700 days. Individual RVs have a range
from $-11$ to $-54$~\ks. \citet{becker2015} published the mean RV of \va
derived from the high-resolution echelle spectrograph 
(HIRES) attached to the Keck~I reflector:
$-1.98\pm11.46$~\ks. They noted that the large rms
of the mean indicates some variability. They calibrated the wavelength
scale of the HIRES spectrograph with the iodine gas cell and derived the RVs
using a rather complicated fit of the observed spectra over a large interval
of wavelengths to a template spectrum chosen from the grid of PHOENIX models
(see Sect.~2 of their paper for details.) \citet{wang2018} searched
for new hot sub-dwarf companions to Be stars in the International Ultraviolet
Explorer (IUE) spectra using the cross-correlation (CCF) technique. Their search for \va
gave marginal detections in only a part of the available IUE spectra and they tabulated
25 RVs of the possible sub-dwarf companion. Later, however, \citet{wang2021}
investigated high-quality far-UV spectra from the Imaging Spectrograph of the Hubble
Space Telescope (HST/STIS), again using the CCF technique,
and were unable to find a trace of a~hot compact secondary.
They only published three RVs of the Be star secured over an interval of 35 days,
all close to $-25$~\ks.
\citet{klement2022}, maybe unaware of the \citet{losh32} study, claim that no evidence
of a~spectroscopic orbit is available in the literature. Using the Gaia distance and
indirect arguments, they estimated the probable orbital period at a value of 246~d,
close to the \citet{losh32} result. \citet{klement2024} obtain 12 good interferometric
observations covering an interval of 457 days and conclude that \va is an astrometric and
spectroscopic binary with a 359\fd26$\pm$0\fd041 period and a~circular orbit.
Their interferometry covers about one half of the 359~d orbit.

Since this latest value of the orbital period is very close to one tropical year, which
is a bit suspect, and considering the general importance of this object,
we decided to carry out a new detailed study based on the numerous spectra available
to us and on published RVs and spectrophotometric measurements.
Our goals are (i) to measure the emission and absorption RVs and to find the true
period of RV changes, and (ii) to document the time variability of the object over a long
time interval. A detailed study of rapid spectral and brightness changes, also based on 
Transiting Exoplanets Survey Satellite (TESS) photometry, will be the subject of another study.

\begin{table}
\begin{center}
\caption[]{Journal of electronic spectra.}\label{jouspe}
\begin{tabular}{ccrcrl}
\hline\hline\noalign{\smallskip}
Spg.&Time interval&No.    &Wavelength &Spectral  \\
 No.&             &of     & range     &res.\\
    &(HJD-2400000)&spectra& (\AA)   \\
\noalign{\smallskip}\hline\noalign{\smallskip}
 1&49947.67--60507.72& 35&6320--6920&20000\\
 2&52844.39--57238.54& 19&6270--6720&12000\\
 3&54393.40--60442.55&102& near \ha &various\\
 4&59331.93--59829.74& 35&6320--6920&20000\\
 5&59429.54--60447.52& 88&4000--7000&53000\\
 6&59436.89--59805.72& 38&4000--7000&53000\\
\noalign{\smallskip}\hline\noalign{\smallskip}
\end{tabular}
\tablefoot{Column `Spg. No.': \ \
1) DAO 1.22 m reflector, coud\'e grating McKellar spectrograph,
     CCD Site4 detector, spectra from programs of SY and PK;
2) Ond\v{r}ejov 2.0 m reflector, coud\'e grating spectrograph,
     CCD detector;
3) A selection of amateur spectra from the BeSS database with
     resolutions better than 11000 (\url{http://basebe.obspm.fr/basebe});
4) DAO 1.22 m reflector, coud\'e grating McKellar spectrograph,
     CCD Site4 detector, spectra from the program of JLB;
5) NRES network, spectra from Wise Observatory (tlv);
6) NRES network, spectra from Mc Donald Observatory (elp).
}
\end{center}
\end{table}

\begin{table}
\begin{center}
\caption[]{Journal of published RVs.}\label{jourv}
\begin{tabular}{crcl}
\hline\hline\noalign{\smallskip}
     Time interval&No.&Instr.&Reference         \\
                  &of &      &                  \\
     (HJD-2400000)&RVs&      &                  \\
\noalign{\smallskip}\hline\noalign{\smallskip}
19219.74--25529.67&205& 11   &\citet{losh32}    \\
41881.77--42602.83& 24& 12   &\citet{abt78}     \\
53233.80--54031.61&  4& 13   &\citet{grundstrom2007}\\
53290.64--54022.60&  4& 14   &\citet{grundstrom2007}\\
55074.06--56135.05& 63& 15   &\citet{becker2015}\\
58718.14--58753.77&  3& 16  &\citet{wang2021}\\
\noalign{\smallskip}\hline\noalign{\smallskip}
\end{tabular}
\tablefoot{Column `Instrument': \ \
11) Ann Arbor 0.95 m reflector, 1-prism spg., photographic \hb, \hg, and H$\delta$ spectra;
12) KPNO 0.9 m reflector, coud\'e feed grating spg., blue photographic spectra;
13) KPNO 0.9 m reflector, coud\'e feed grating spg., blue CCD T2KB spectra;
14) KPNO 0.9 m reflector, coud\'e feed grating spg., red CCD T1KB spectra;
15) KECK~I 10 m reflector, HIRES echelle CCD spectrograph with spectral resolution of 55000;
16) HST far UV spectra, RVs measured by cross-correlation technique.
}
\end{center}
\end{table}

\section{Observations and data reductions}
\subsection{Spectroscopy}
We carefully reduced numerous red electronic spectra from several observatories,
also including a selection of BeSS amateur spectra \citep{neiner2011}, which had
spectral resolutions better than 11000, and we measured the \ha and \he RVs
using the latest version~2 of the program
\respefo\footnote{https://astro.troja.mff.cuni.cz/projects/respefo/} and comparing
the direct and flipped line profiles on the computer screen.
To have some idea of the uncertainties in the settings on the lines, which were 
affected by either the presence of numerous telluric lines or noise, all RVs were 
independently measured twice by Petr~Harmanec and Pavel~Dole\v{z}al, and the mean 
values were adopted. A record of all the spectra used is provided in Table~\ref{jouspe}.

Additionally, we collected (and digitized when necessary) the existing published
RVs with known times of observation. An overview of this is provided in Table~\ref{jourv}.
All individual RVs and their heliocentric Julian dates (HJDs) adopted from 
the literature are in Table~3, and the RVs measured in the electronic spectra by us are 
in Table 4. Both these tables are only available at the Strasbourg astronomical data
centre (CDS).

\subsection{Photometry}
In the past, \va was monitored in a carefully calibrated stardard \ubv\ system
at Hvar, but regrettably this has not happened in recent years. 
We resumed observations in 2024, using the \ubvr\ photometry.  
Additionally, we transformed the existing
Hipparcos \hp\ observations with flags 0 and 1 to Johnson $V$ magnitude
after \citet{hpvb}, and we compiled existing photometric observations with known dates
of observation, which had been obtained in the Johnson \ubv\ system or could 
be transformed to it. They are summarized in Table~\ref{jouphot} and published individually
in Table~6. This table is also only available at the CDS.

\setcounter{table}{4}
\begin{table*}
\caption[]{Journal of available photometry with known dates of observation.}
\label{jouphot}
\begin{flushleft}
\begin{tabular}{rcrccll}
\hline\noalign{\smallskip}
Station&Time interval& No. of &Passbands&HD of comparison&Source\\
       &(HJD$-$2400000)&obs.  &  &/ check star\\
\noalign{\smallskip}\hline
\hline\noalign{\smallskip}
23&38186.95--38305.64&  4&\ubvri      & all-sky     & \citet{john66}\\
01&42237.49--43015.42& 25&\ubv        & all-sky     & This paper\\
01&46235.62--47791.36&304&\ubv        &188892/193369& This paper\\
01&60468.51--60583.32&138&\ubv        &188892/193369& This paper\\
20&44756.81--49951.77&110&$BV$        &188892/193369& \citet{percy97} and priv.com.\\
61&47892.69--49046.08&145&\hp-->$V$   &all-sky      & \citet{esa97}\\
15&48041.89--48598.58&155&\ubv        &188892/193369& \citet{percy97} and priv.com.\\
36&48779.96--48801.93&  6&\uvby-->\ubv&188892/193369& \citet{adelman97}  \\
66&52767.49--52768.55&  6&\ubv&188892/193369& This paper \\
\noalign{\smallskip}\hline
\end{tabular}
\tablefoot{In the column {\sl `Station',} the individual observing stations are
identified by the running numbers from the Praha/Zagreb photometric archives:
01) Hvar 0.65~m reflector, EMI tubes;
15) Phoenix 10 APT 0.25~m reflector \citep{seeds92};
20) Toronto 0.40~m reflector, EMI 6094 tube;
23) Catalina Station of Lunar and Planetary Observatory, 0.54~m and 0.70~m reflectors,
      1P21 photomultiplier;
36) Four College 0.75 m APT, Mt. Hopkins, Stromgr\"en \uvby\ filters transformed to \ubv\
      after \citet{hecboz2001};
61) Hipparcos \hp\ photometry transformed into Johnson $V$ after \citet{hpvb};
66) T\"ubitak National Observatory 0.40~m reflector, SSP-5A photometer.}
\end{flushleft}
\end{table*}

\section{Mapping the timescales of variations}
\begin{figure}[t]
\resizebox{\hsize}{!}{\includegraphics[angle=0]{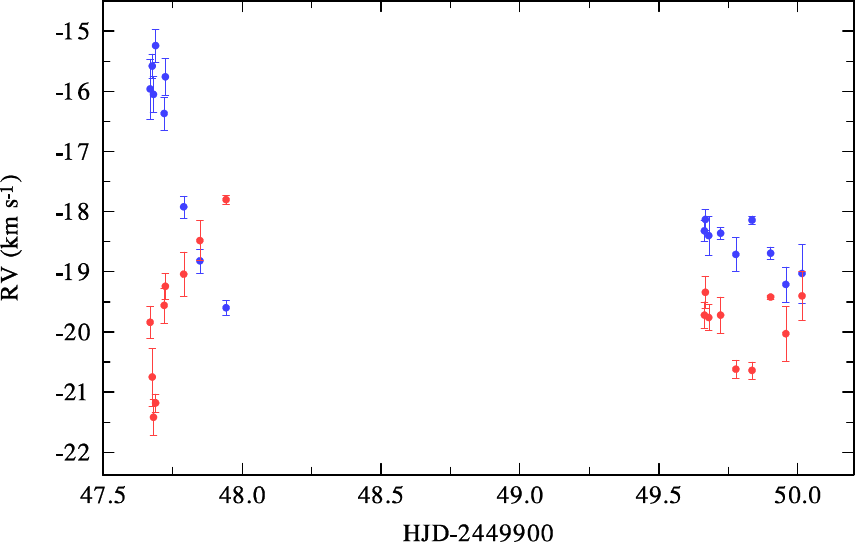}}
\resizebox{\hsize}{!}{\includegraphics[angle=0]{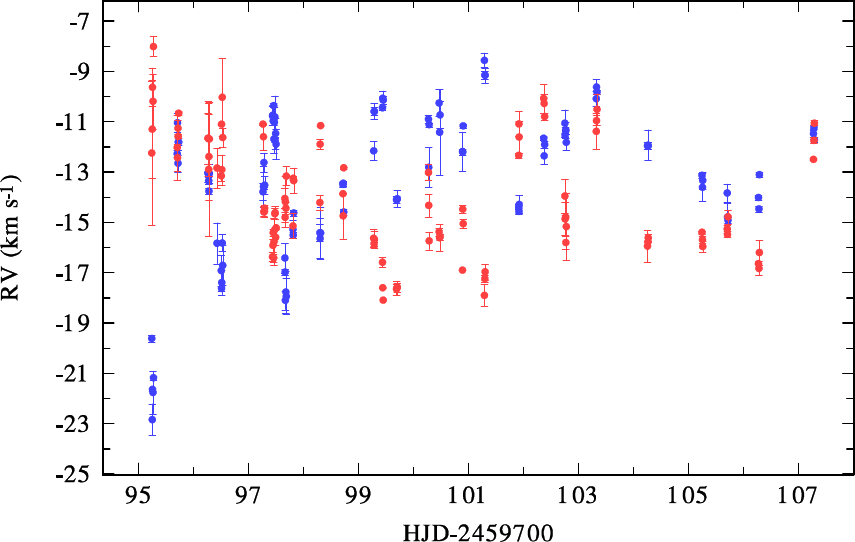}}
\caption{Evidence of rapid RV variations in the \ha line measured in two
dense night series of spectra. RVs of the emission wings are denoted
by red circles, and those of the absorption core by blue circles.
The error bars of individual RVs are shown.
Top: A series of DAO spectra. Bottom: A series of NRES spectra.
}
\label{rapidrv}
\end{figure}

First we wanted to see how the known rapid spectral and brightness variations
affect our RV measurements of the steep emission wings and absortion core
of \hae. To this end, we selected two dense series of observations, from 
the Dominion Astrophysical Observatory (DAO) and the network of robotic echelle
spectrographs (NRES) and plotted the measured RVs versus time, see Fig.~\ref{rapidrv}.
One can see that the RVs are indeed affected by rapid changes and there is some
tendency to anti-correlation between emission and absorption RVs.

\begin{figure}[t]
\resizebox{\hsize}{!}{\includegraphics[angle=0]{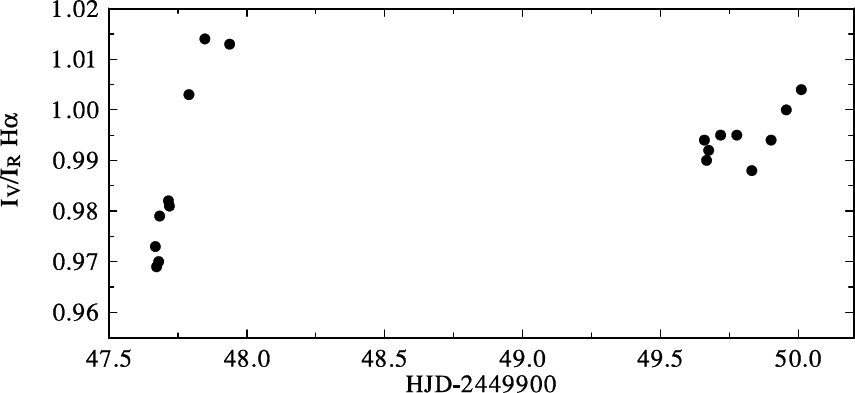}}
\resizebox{\hsize}{!}{\includegraphics[angle=0]{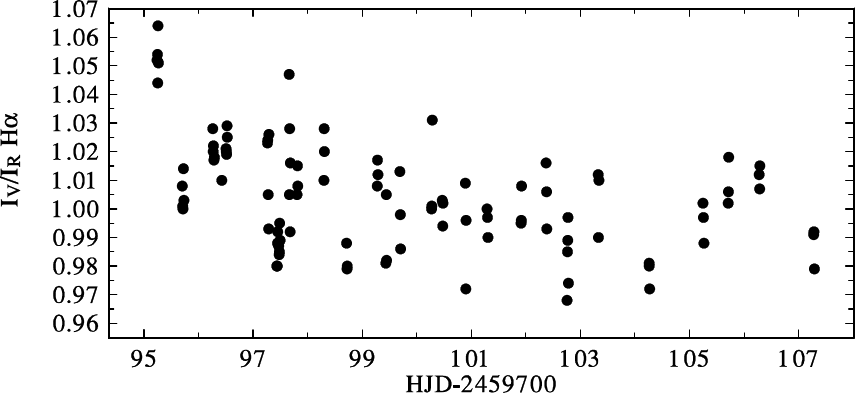}}
\caption{Evidence of rapid $V/R$ variations in the \ha line measured in two
dense night series of spectra.  Top: A series of DAO spectra.
Bottom: A series of NRES spectra.
}
\label{rapidvr}
\end{figure}

In Fig.~\ref{rapidvr} we show the $V/R$ changes of the double \ha emission
for the same two night series of spectra. It seems probable that the
measured RVs were partly affected by these line-profile changes.
We note that similar rapid changes were reported by \citet{bossi93},
who reconciled them with a~period of 0\fd7564.
We also note that the NRES cadence was high specifically because
the rapid variations of the \ha emission were observed during an~active mass
ejection phase studied by \citet{bartz2025}.
One of the main conclusions of their paper is that, in all observed cases,
Be emission lines always exhibit rapid asymmetry oscillations with
a~characteristic frequency during and/or shortly after the mass ejection
episodes, which would likewise also manifest in RV variations when using
the emission wing mirror method.

\section{Long-term variability of \ve}
When we compiled the available spectrophotometric and photometric observations
of \ve, it became evident that \va has been undergoing irregular long-term
cyclic changes in several observables.
This finding is in agreement with previous studies.

\begin{figure}[t]
\resizebox{\hsize}{!}{\includegraphics[angle=0]{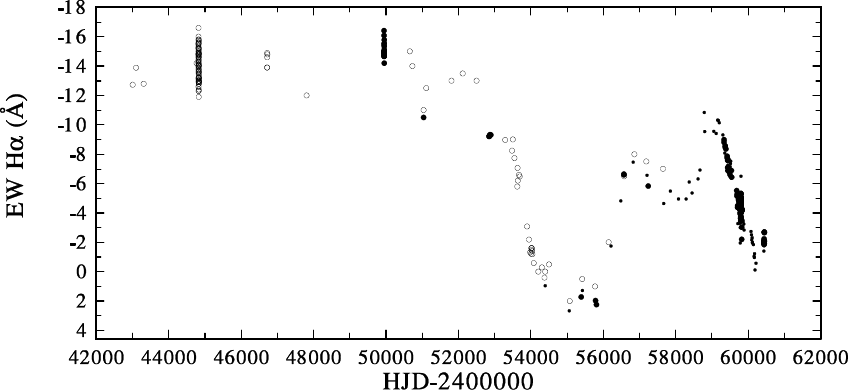}}
\resizebox{\hsize}{!}{\includegraphics[angle=0]{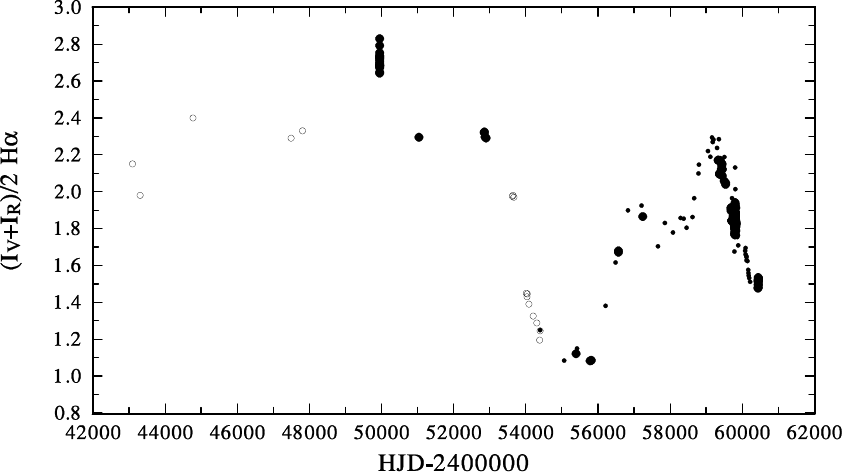}}
\caption{Recorded secular changes of the equivalent width (upper panel)
and the peak intensity (bottom panel) of the \ha line.
The black dots are our measurements in electronic spectra, smaller symbols
denote the BeSS amateur spectra, open circles are measurements adopted
from the papers of \citet{slet78,fontaine82,andril83,tolja86,doazan91,slet92,
bossi93,grundstrom2007,hessel2009,jones2011}, and \citet{baade2018}.
}
\label{h3ew}
\end{figure}

\begin{figure}[t]
\resizebox{\hsize}{!}{\includegraphics[angle=0]{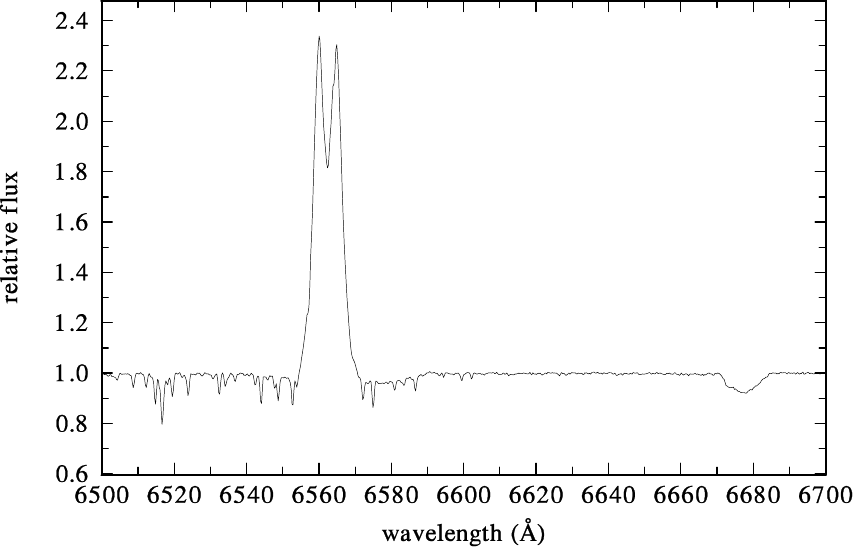}}
\resizebox{\hsize}{!}{\includegraphics[angle=0]{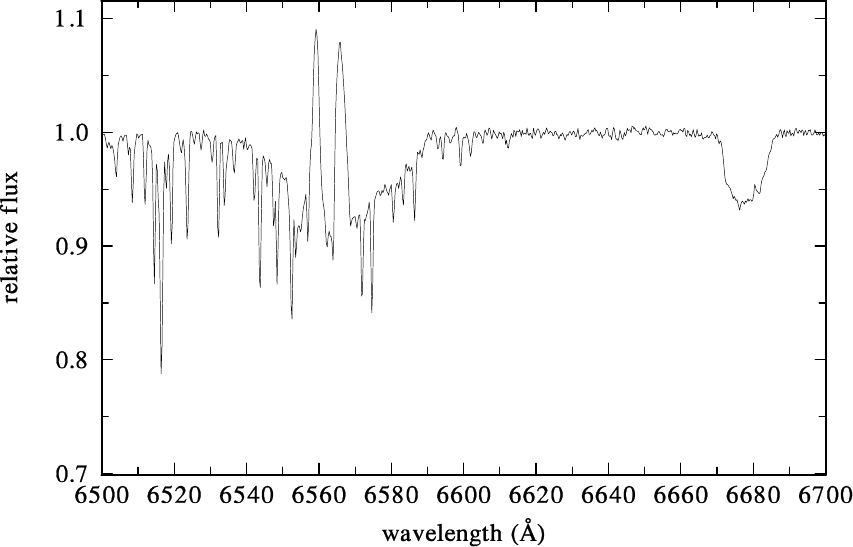}}
\caption{Parts of the Ond\v{r}ejov red spectra taken on
HJD~2452844.3888, when the \ha emission was close to its maximum strength       
(top panel), and on HJD~2455814.3065, i.e. near the deepest minimum of
all recorded equivalent width measurements (bottom). We see that
the double \ha emission is still clearly present there.
}
\label{profil}
\end{figure}

Figure~\ref{h3ew} shows the dramatic changes in the equivalent width
of the \ha line with time. In contrast to some other Be stars, it seems
that \va has never been observed in a purely absorption phase. 
\citet{losh32} discussed evidence for an episode of long-term $V/R$
and line intensity changes of the Balmer lines for Julian dates (JDs)
prior to 2420000. Even near the equivalent-width minimum around 
JD~2455000-2456000, a~double \ha emission was clearly present 
(see the bottom panel of Fig.~\ref{profil}).

\begin{figure}[t]
\resizebox{\hsize}{!}{\includegraphics[angle=0]{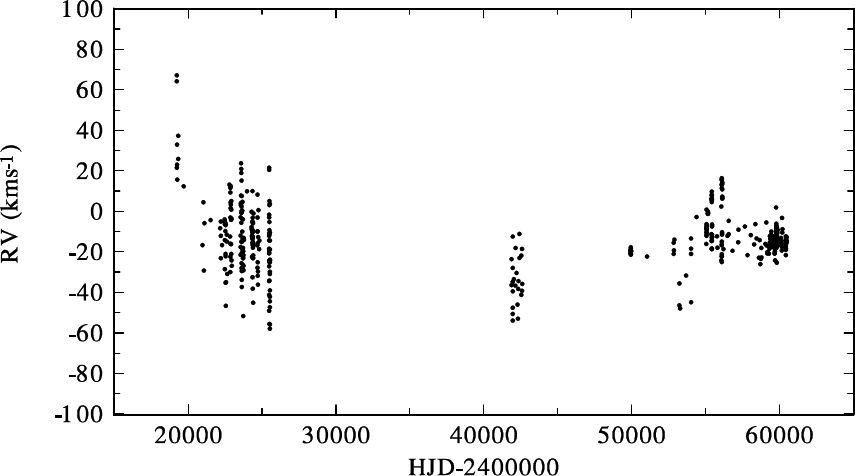}}
\resizebox{\hsize}{!}{\includegraphics[angle=0]{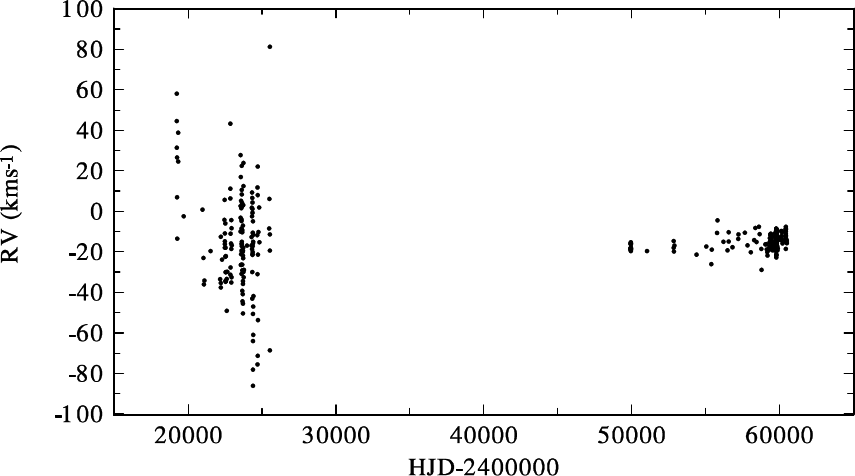}}
\caption{Time plot of all RVs at our disposal. Absorption RVs are
shown in the upper plot, RVs of \ha emission wings are shown in the
bottom plot.
}
\label{rvtime}
\end{figure}

In Fig.~\ref{rvtime} we show a plot of all RVs for the data at our disposal,
separate for the absoption RVs, and for Balmer emission RVs. Here,
the evidence of secular changes is less obvious and can only be suspected
for the observations before JD~2421000, as already noted by \citet{losh32}.
It must also be kept in mind that the available material is somewhat
heterogenous. \cite{losh32} RVs are based on the mean of Balmer \hb
and \hg RVs, the RVs by \citet{abt78} are mean RVs of a number of Balmer and
\ion{He}{I} from photographic spectra (RVs around JD~2442000), the RVs by
\citet{grundstrom2007} are tabulated for Balmer line cores and \ion{He}{i}
lines, the RVs from the Keck~I telescope are cross-correlation RVs, and all our
RVs are measured from \ha line profiles.

\begin{figure}[t]
\resizebox{\hsize}{!}{\includegraphics[angle=0]{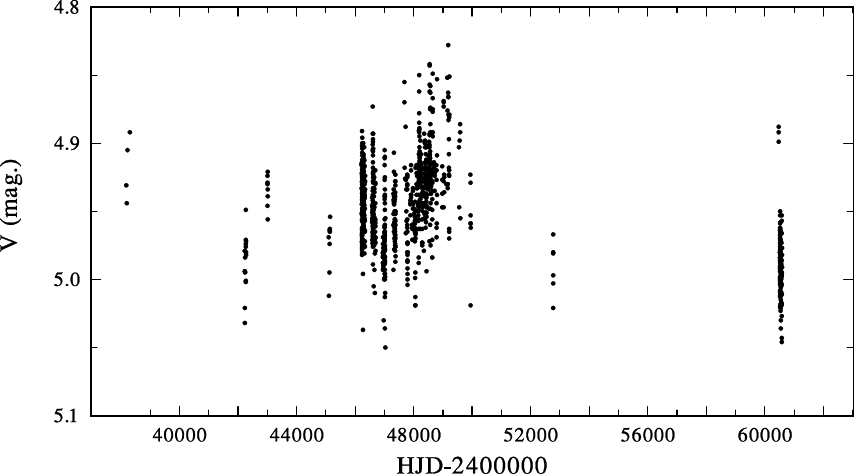}}
\resizebox{\hsize}{!}{\includegraphics[angle=0]{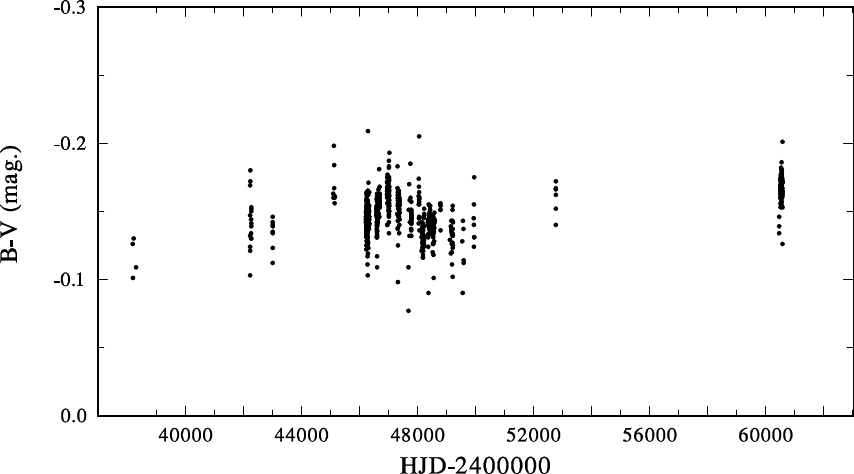}}
\resizebox{\hsize}{!}{\includegraphics[angle=0]{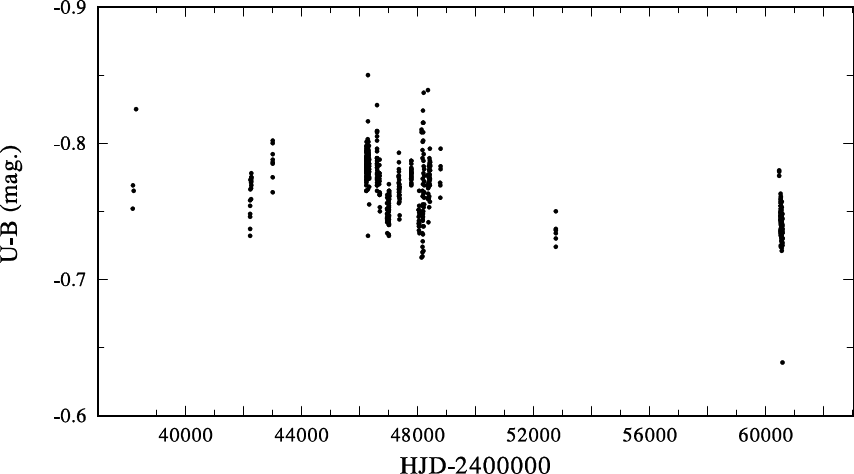}}
\caption{Recorded brightness and colour changes based
on observations secured in, or transformed to,
the Johnson \ubv\ system.
}
\label{ubvall}
\end{figure}

Finally, in Fig.~\ref{ubvall} we plot the available records of the brightness and colour
changes based on observations obtained in, or transformed to, the Johnson \ubv\ system.
Besides obvious rapid changes, one can see mild secular changes, possibly correlated
with the long-term spectral variations. Regrettably, observations from the interval of
of JD~2454000-55000, when the Balmer emission was weak, are missing.

\section{Towards orbital solution}
\begin{figure}[t]
\resizebox{\hsize}{!}{\includegraphics[angle=0]{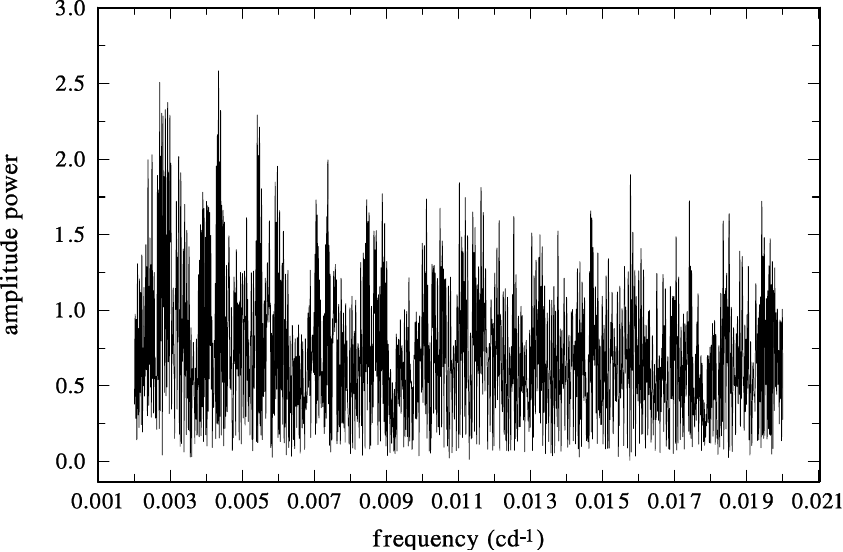}}
\resizebox{\hsize}{!}{\includegraphics[angle=0]{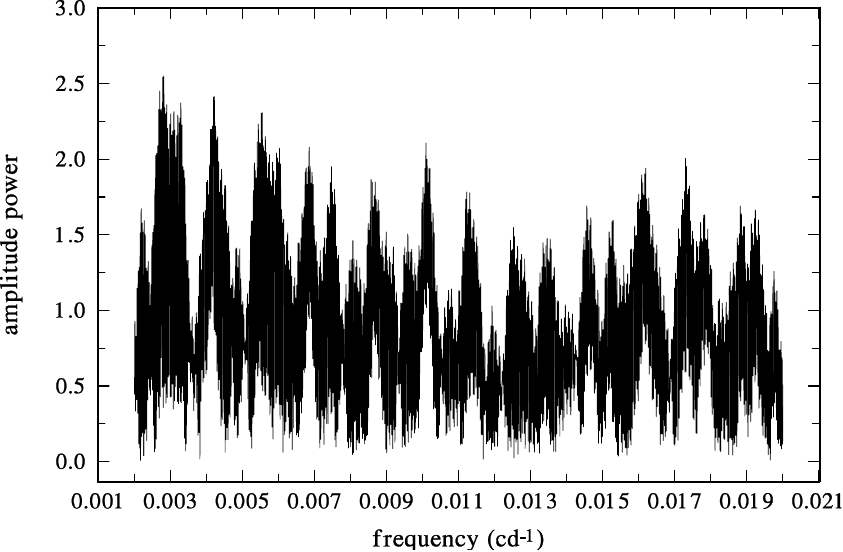}}
\caption{Amplitude periodograms of all RVs without early data from \citet{losh32}.
Top: Absorption RVs. Bottom: \ha emission-line wings RVs.
The frequency of the orbital period of 359\fd26 is
0.002783~\cd, its harmonics is 0.005567~\cd, and the frequency of
the 236 d period is 0.004237~\cd.
}
\label{power}
\end{figure}

It is our experience that the RVs measured on the steep emission-line wings
(see two samples of the \ha line profiles in Fig.~\ref{profil}) usually provide 
the most realistic orbital RV curve for the Be components of binary stars
\citep[see the detailed justification in][]{zarf26}.
Having a compilation of such emission-line velocities and also a larger but rather
heterogeneous set of absorption RVs, we subjected both sets to period analyses
with a program based on the \citet{deeming75} method, which derives amplitude
periodograms. In doing so, we excluded early Ann Arbor RVs secured before JD 2420000,
which are affected by long-term changes
\citep[see the discussion in][and Fig.~\ref{rvtime} here]{losh32}.
Since the shortest period identified by \citet{baade2018} in several sets of
photometric observations was about 0\fd31, we carried out the period searches in
the interval from 0\fd31 to 500\fd0, with a frequency step of 0.1/$\tria$T,
where $\tria$T denotes the time interval covered by observations. Besides
the one-day and two-day aliases of the long period,
no significant rapid periods were detected. For both datasets, the two highest peaks
in the periodograms were close to 355-360 d and 236 d. For the emission-line RVs,
the best period was very close to the 359\fd26 period found by \citet{klement2024}
from interferometry. The amplitude periodograms are shown in Fig.~\ref{power}.

\setcounter{table}{6}
\begin{table}
\begin{center}
\caption[]{Ann Arbor emission-line RVs with the \oc\ residuals larger than
40 \ks, omitted from the final orbital solution.}\label{omit}
\begin{tabular}{cccc}
\hline\hline\noalign{\smallskip}
HJD-   &No.&HJD-     &No.\\
2400000&   &2400000\\
\noalign{\smallskip}\hline\noalign{\smallskip}
22846.720&1&24386.700&1\\
23620.747&1&24679.788&1\\
23736.623&2&24697.723&1\\
24373.701&1&24701.809&1\\
24379.778&1&25522.584&1\\
24385.710&1&25525.621&1\\
\noalign{\smallskip}\hline\noalign{\smallskip}
\end{tabular}
\end{center}
\end{table}

\begin{figure}[t]
\resizebox{\hsize}{!}{\includegraphics[angle=0]{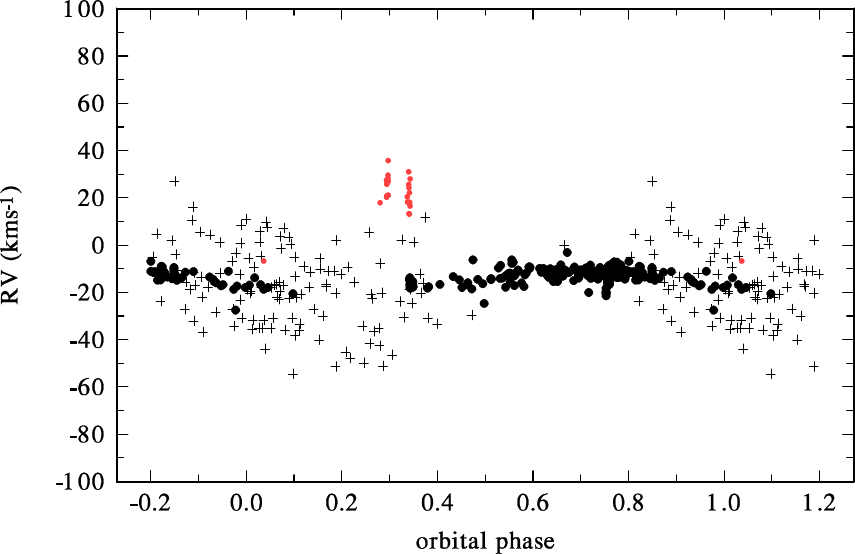}}
\resizebox{\hsize}{!}{\includegraphics[angle=0]{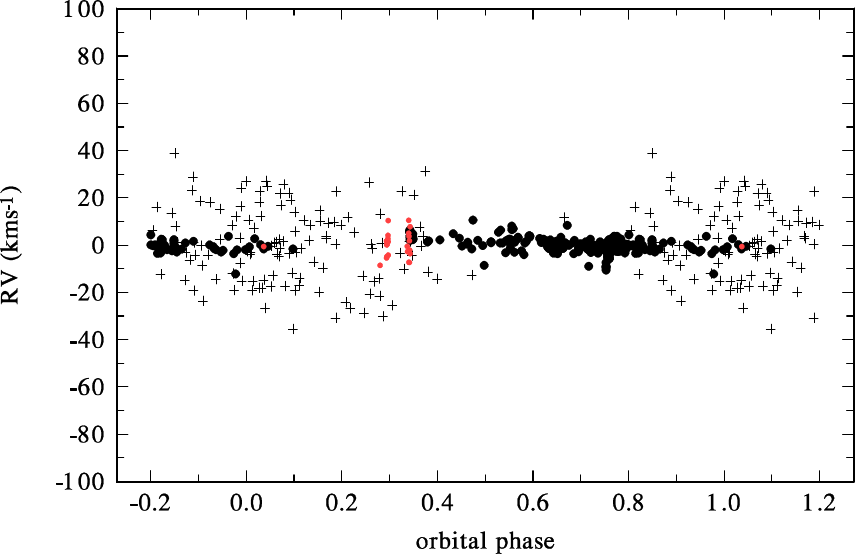}}
\caption{Radial-velocity curve for the final solution (top panel)
and the \oc\ residuals from it (bottom panel). RVs are emission-line
RVs, the mean of \hb\ and \hg\ from the Ann Arbor photographic spectra by
\citet{losh32} (shown by pluses), and \ha\ emission wings for all
CCD spectra (shown as black circles). The IUE spectra of the compact
secondary, published by \citet{wang2018}, are shown by red circles.
}
\label{rvcurve}
\end{figure}

\begin{table}
\begin{flushleft}
\caption{Orbital solutions based on the \ha emission RVs.}
\label{sol}
\begin{tabular}{lccrrccc}
\hline\hline\noalign{\smallskip}
 Element                   & Solution 1    & Solution 2         \\
\noalign{\smallskip}\hline\noalign{\smallskip}
$P$ (d)                    & 359.26 fixed  &$358.98\pm0.27$     \\
$T_{\rm super.conj.} ^*$   & 59524.8 fixed &$59509.9\pm5.9$     \\
 $e$                       &0.0 fixed      & 0.0 fixed          \\
$\gamma_{\rm CCD}$ (\ks)   &$-17.34\pm0.58$&$-17.53\pm0.61$     \\
$\gamma_{\rm phg}$ (\ks)   &$-15.2\pm1.4$  &$-15.8\pm1.8$       \\
$\gamma_{\rm IUE}$ (\ks)   &$-25.7\pm6.2$  &$-26.5\pm7.8$       \\
$K_1$    (\ks)             & $5.10\pm0.88$ & $5.40\pm0.87$      \\
$K_1/K_2$                  &$0.118\pm0.033$&$0.117\pm0.038$     \\
$K_2$    (\ks)             & 43.33         & 46.16              \\
$m_1\sin^3i$ (\ms)         & 3.78          & 4.57               \\
$m_2\sin^3i$ (\ms)         & 0.45          & 0.53               \\
$a\,\sin i$ (\rs)          & 343.9         & 365.9              \\
No. of RVs (prim./sec.)    & 432/25        & 432/25             \\
rms$_{\rm CCD}$ (\ks)      &  2.97         &  2.88              \\
rms$_{\rm phg}$ (\ks)      & 15.75         & 15.71              \\
rms$_{\rm IUE}$ (\ks)      &  5.05         &  5.25              \\
\hline\noalign{\smallskip}
\end{tabular}
\end{flushleft}
\tablefoot{$^*$) All epochs are in HJD-2400000.}
\end{table}

We then derived several trial orbital solutions using program \fotel
\citep{fotel1, fotel2}. For the emission RVs, this led to reasonable
solutions that supported the orbit derived by \citet{klement2024}. However,
it turned out that for some of the Ann Arbor photographic RVs, based on
\hb and \hg emission-wing velocities, the \oc\ residuals were quite high.
We suspect that this happened in cases when one of the emission peaks
of the double emission lines was stronger and the setting was made on it
instead of on the wings of the whole line. As a precaution, we omitted 12
such RVs, which had residuals larger than 40~\ks, from the solution.
In all cases, with one exception, these measurements were based on
only one Balmer line. We list the JDs of these omitted RVs
in Table~\ref{omit}. We then calculated two other orbital solutions:
solution 1, in which the orbital period and epoch of superior
conjunction were fixed at values from the accurate interferometric
solution by \citet{klement2024}, and solution 2, where we allowed
for the convergency of these elements. To obtain the estimate of
the mass ratio as well, we included 25 RVs from the IUE, reconstructed
by \citet{wang2018}. Both these solutions are listed in Table~\ref{sol}.
We underline that a~solution based on all Ann Arbor spectra has
naturally larger rms errors, but the elements are quite similar
to those of solution~1. We note that we allowed for different systemic
velocities for the CCD, IUE, and Ann Arbor photographic spectra since
they can have different zero points of the velocity scale. All CCD
spectra were assigned the same systemic velocity since their RV zero
point was checked through the RV measurements of a~selection of telluric
lines. It is encouraging that the systemic velocities of CCD and
Ann Arbor spectra agree within their estimated errors. This increases
the credibility of the solutions. As Fig.~\ref{rvcurve} shows, available
RVs from electronic spectra do not cover the RV minimum and
the orbital solution based solely on them leads to a significantly lower
semi-amplitude of the velocity curve. This is the reason why we currently
prefer the solutions based on all available emission-line RVs. For a final,
fully reliable orbit of the Be primary, continuing spectral observations
in phases of strong Balmer \ha emission are needed.
Finally, we mention that a solution
for all absorption-line RVs leads to an orbital period and epoch that are
quite similar to those of solution~2.

We also tried to recompute the elements of the interferometric orbit
using an independent program for such a~solution \citep{zasche2007}.
Given the limited number of only 12 observations, we arrived at very 
similar solutions to those presented in Table~\ref{sol}. Only with a~higher number
of precise interferometric observations would it be possible to discriminate
between the two solutions of Table~\ref{sol}.

\begin{figure}[t]
\resizebox{\hsize}{!}{\includegraphics[angle=0]{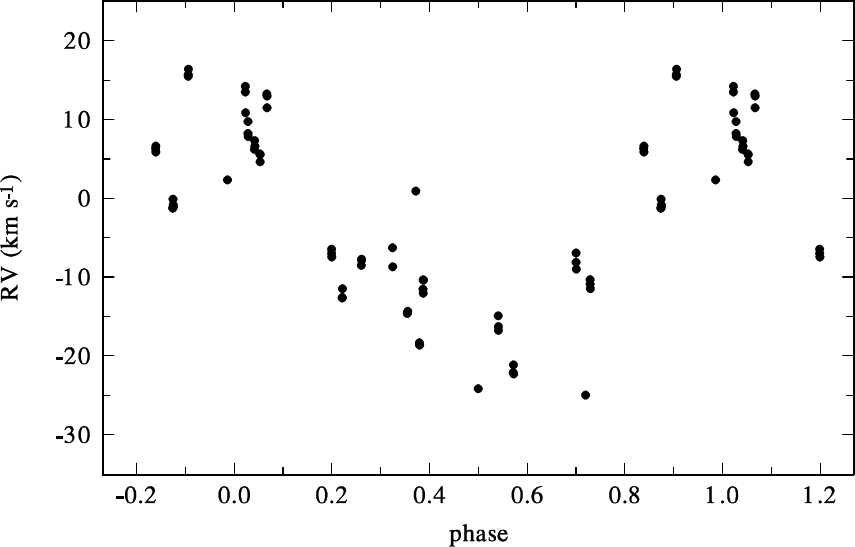}}
\caption{Phase diagram of Keck RVs plotted for the ephemeris
$T_{\rm RVmax.}={\rm HJD}~2455501.016(22)+1\fd84290(12)\cdot E$.
}
\label{keck}
\end{figure}

At the suggestion of an anonymous referee, we tried to check whether
a~reasonable orbital solution could be obtained for absorption RVs from
a~subset of electronic spectra with high spectral resolutions, i.e. NRES
and KECK spectra. The trouble is that while the NRES spectra, based on
the measurements of the \ha absorption core in \respefoe, cover only a limited
phase interval of the 359~d orbital period, the KECK RVs, based on a comparison
of observed spectra with the synthetic spectra over a large interval of
wavelengths, are dominated by rapid RV changes and do not give a reasonable
RV curve for the orbital period. A period search in these RVs and
a consecutive sinusoidal fit returned a period of 1\fd84290(12), which is close
to a~one-day alias of the period of 0\fd6474 (frequency 1.5447~\cd), one of
the dominant periods found by \citet{baade2018} in their period analyses
of several satellite photometries of \ve. A~corresponding phase plot is shown in
Figure~\ref{keck}. No orbital signal is obvious, even in the residuals from
the fit of the 1\fd84 period.

\section{Discussion}
   When we provisionally take solution~1 as the final one, 
and assume an orbital inclination
$i=61.^\circ3$ after \citet{klement2024}, we obtain the binary masses of
$m_1=5.6$~\ms\  and $m_2=0.66$~\ms. \citet{zorec2016} attempted to estimate
the fundamental parameters of a group of Be stars. For \va, they obtained
\tef = $23830\pm930$~K and the inclination of the rotatonal axis estimated
from the gravity darkening as $69^\circ\pm17^\circ$. Within the limits of
the respective errors, it seems that the orbital and rotational axes are aligned.
The above value of \tef\ would correspond to a primary mass of
about 9~\ms\ according to \citet{mr88}. On the other hand, the Be primary was
often classified as B3V and the mass of 5.6~\ms\ is then acceptable according
to tabulation by \citet{mr88}. The total mass we obtained is slightly lower than
that obtained by \citet{klement2024}. These authors derived the total mass using
the IUE RVs of the secondary from \citet{wang2018} and the Gaia distance, which,
however, was not yet corrected for the effect of the 359~d orbit. Moreover,
if one takes the estimated errors of $K_1$ and $K_1/K_2$ of our solution~1
into account, the two results agree within the error limits. To obtain
an~accurate determination of both binary masses, not only continuing \ha
spectra of the primary but also new far-UV spectra of the secondary will
be needed. It is not quite clear why \citet{wang2021} were unable to detect
the lines of the secondary in the good HST/STIS far-UV spectra taken near
one of the elongations. They tentatively suggested that some kind of
temporal variability makes the detection of the secondary more favourable
at certain epochs only.

  We note that the orbital period of \va is one of the longest
orbital periods yet confirmed for a~Be binary with a~post-mass-transfer
companion. The inspection of a catalogue of binaries with hot components
and Balmer emission lines by \citet{hec2001} shows that only rather
exotic objects like symbiotic stars have longer orbital periods. One
system that bears some similarity is KS Per = HD 30353, with a 362\fd8 period
and a hot sub-dwarf companion.

This underscores the importance of long-term monitoring of Be stars.
With such a small semi-amplitude of only $\sim5$ \ks, and Porb $\sim1$~yr,
confirming binaries such as \ve, and perhaps any yet-undiscovered even 
longer-period Be binaries, requires long time series of high-quality data and careful
analysis. It is likewise important to inspect older literature records. These
often contain valuable insights, even if the at-the-time interpretation has
changed. That we are only recently coming to understand the binary properties of
such a~bright and well-studied star as \va suggests that we are still
exploring the 'tip of the iceberg' of the Be population.

\begin{acknowledgements}
We gratefully acknowledge the use of the latest publicly available version
of the program \fotele, written by P.~Hadrava. We also acknowledge the use
of the reduction program for spectroscopic reductions \respefoe,
written by A.~Harmanec. Our sincere thanks
go to J.C.~Becker and A.~Vanderburg, who kindly provided us with their individual
RV measurements of \va from the Keck~I telescope spectra, and to J.R.~Percy,
who put his photoelectric observations at our disposal. Recommendations by 
an anonymous referee improved the clarity of this study.
Over the years, this long-term project was supported by the grants 205/06/0304,
205/08/H005, P209/10/0715, and GA15-02112S  of the Czech Science Foundation,
by the grants 678212 and 250015 of the Grant Agency of the Charles University
in Prague, from the research project AV0Z10030501 of the Academy of Sciences
of the Czech Republic, and from the Research Program MSM0021620860
{\sl Physical study of objects and processes in the solar system and
in astrophysics} of the Ministry of Education of the Czech Republic.
The research of PK was supported by the ESA PECS grant 98058.
HB acknowledges financial support
from the Croatian Science Foundation
under the project 6212 ``Solar and Stellar Variability".
We used some spectra of the BeSS database, operated at LESIA,
Observatoire de Meudon, France: \url{http://basebe.obspm.fr}.
Finally, we acknowledge the use of the electronic database from
the CDS, Strasbourg, and the electronic bibliography maintained by
the NASA/ADS system.
\end{acknowledgements}

\bibliographystyle{aa}
\bibliography{zarfin}

\end{document}